\documentclass[journal]{IEEEtran}

\usepackage{graphicx}  

\begin{document}

\title{Experimental Information and Statistical Modeling of Physical Laws}

\author{Igor~Grabec
\thanks{Manuscript received March, 2006; revised:}
\thanks{I. Grabec is with the Faculty of Mechanical Engineering, University of Ljubljana, Slovenia (e-mail:igor.grabec@fs.uni-lj.si).}}

\markboth{IEEE Transactions on Information Theory,~Vol.~?, No.~??,~Month?~2006}{Grabec: Experimental Information and Statistical Modeling of Physical Laws}

\maketitle

\begin{abstract}
Statistical modeling of physical laws connects experiments with mathematical descriptions of natural phenomena. The modeling is based on the probability density of measured variables expressed by experimental data via a kernel estimator. As an objective kernel the scattering function determined by calibration of the instrument is introduced. This function provides for a new definition of experimental information and redundancy of experimentation in terms of information entropy. The redundancy increases with the number of experiments, while the experimental information converges to a value that describes the complexity of the data. The difference between the redundancy and the experimental information is proposed as the model cost function. From its minimum, a proper number of data in the model is estimated. As an optimal, nonparametric estimator of the relation between measured variables the conditional average extracted from the kernel estimator is proposed. The modeling is demonstrated on noisy chaotic data.
\end{abstract}

\begin{keywords}
kernel estimator, experimental information, complexity, redundancy, modeling of a physical law, model cost function, conditional average predictor, nonparemtric regression, predictor quality, noisy chaotic generator
\end{keywords}

\IEEEpeerreviewmaketitle

\section{Introduction}
\PARstart{E}{xperimental} exploration of natural phenomena includes measurements and descriptions of corresponding physical laws \cite{fe}.
Modern experimental systems can perform measurements automatically, and therefore the question arises of how to develop a system for an automatic description of physical laws \cite{gs}. Tools involved in both tasks of exploration differ essentially in their character: measurements are based on devices and provide data about measured variables, while descriptions are based on mathematical methods and yield relations between these variables \cite{fe}. Nature has a tremendous variety of properties, which results in the diversity of functions applicable to mathematical modeling of corresponding physical laws. On the contrary, a unique function is needed for an automatic modeling of physical laws in experimental systems. To bridge this gap, we employ the probability distribution \cite{re} as a common basis for the description of natural properties and propose a nonparametric regression as a general method for the experimental modeling of physical laws. The corresponding statistical estimator is the conditional average (CA), which can be automatically extracted from the probability density function (PDF) in a measurement system \cite{gs}.
 
For a nonparametric expression of the PDF, Parzen has proposed a kernel estimator \cite{par,dh}, and his method has been successfully applied to the statistical description of various natural laws governing complex phenomena in a variety of fields \cite{gs,tgp,mgg}. However, a common weakness of these applications is the lack of an objective kernel function and a heuristic selection of the number of data representing the model. The same weakness is characteristic of several heuristic methods developed from Parzen's estimator in the fields of neural networks and artificial intelligence \cite{ha,ka}. Here we avoid this deficiency by specifying the kernel more objectively based upon calibration of the measurement system \cite{par,dh}. This specification requires a statistical description of instrument output scattering during calibration, which further provides for definitions of the indeterminacy of measurements, experimental information, complexity of data, redundancy of measurements, information cost function and estimation of a proper number of data for modeling \cite{re,les,ig,ris,ris2,ct,kol}. 

\section{Estimation of probability distribution}
\label{sec:1}
In order to introduce an objective kernel function we consider a phenomenon that can be explored experimentally by a setup containing only two sensors, since the generalization to more complex cases with several sensors is straightforward. The signals from the sensors are represented by the couple ${\bf z}=(x,y)$. We assume that the phenomenon can be characterized statistically by repetition of measurements yielding sample points in the span of the instrument $S_{\bf z}$. This span is a Cartesian product $S_x \otimes S_y$ of spans corresponding to both channels. We assume that both spans are equal and given by the interval $(-L,L)$.
 
Measurements are generally subject to stochastic disturbances or noise, which makes their outcomes uncertain \cite{les}. The uncertainty is usually represented just by the standard deviation of variables during calibration \cite{gs,les}. However, this is not sufficient to answer the following basic questions: 
\begin{enumerate}
\item How much information can be provided by measurements that are influenced by noise \cite{les,zu}?
\item How many experiments are needed for modeling a physical law corresponding to the phenomenon \cite{gs,ig}?
\item How complex should the model of this law be \cite{ris,ris2,ig,zu,be}?
\end{enumerate}
In the following we try to answer these questions based on information theory. With this aim we first describe the signal scattering during calibration of the instrument and then proceed to the uncertainty of experimental observation. 

For a simultaneous calibration of both instrument channels we have to perform a measurement on an object representing two physical units $u_x$ and $u_y$ which we together denote by the joint unit ${\bf u}=(u_x,u_y)$. The scattering of instrument outputs during calibration is characterized by the joint PDF $\psi({\bf z}|{\bf u})$, which we call the scattering function (SF) \cite{ig,gs,les}. When the interaction between both channels is negligible, the SF is given by the product $\psi({\bf z}|{\bf u})=\psi(x|u_x)\psi(y|u_y)$. Without loss of generality we further consider a case with equal sensors which are subject to mutually independent random disturbances that do not depend on ${\bf u}$. In such cases probability theory suggests expressing the SF as $\psi({\bf z}-{\bf u})={\rm g}(x-u_x,\sigma){\rm g}(y-u_y,\sigma)$, where the Gaussian function
\begin{equation}
{\rm g}(x-u_x,\sigma)\,=
\frac{1}{\sqrt{2\pi}\,\sigma}\exp \biggl[-\frac{(x-u_x)^2}{2\sigma}\biggr]
\end{equation}
describes the scattering of signal $x$. The parameters $u_x$, $\sigma$ represent the mean value and standard deviation of this signal at the calibration and can be statistically estimated. 

When we perform a single measurement we get a sample ${\bf z}_1=(x_1,y_1)$ that represents the mean value of ${\bf z}$ during measurement and, therefore, we express the PDF as $\psi({\bf z}-{\bf z}_1)=\psi(x-x_1)\psi(y-y_1)$. When we repeat the measurements $N$ times we get samples ${\bf z}_i ,\,1\le i \le N$, with which we model the joint PDF by the statistical average:
\begin{equation}
f({\bf z})\,=\,\frac{1}{N}\,\sum_{i=1}^N \psi ({\bf z}-{\bf z}_i) .
\label{pdfxy}
\end{equation}

Properties of particular variables $x,y$ are described by the marginal PDFs $f(x), f(y)$. They are obtained from the joint PDF by integration with respect to one component, as for example:
\begin{equation}
f(x)\,=\,\int_{S_y} f({\bf z}) dy\,=\,\frac{1}{N}\,\sum_{i=1}^N \psi (x-x_i).
\label{pdfxe}
\end{equation}
For the modeling of natural laws the most important is the conditional PDF of the variable $y$ at a given value of $x$, defined as: 
\begin{equation}\label{cpdfe}
f(y|x)\,=\, \frac{f({\bf z})}{f(x)}\,=\,\frac{\sum_{i=1}^N \psi ({\bf z}-{\bf z}_i) }{\sum_{j=1}^N \psi (x-x_j) }
\end{equation}

\section{Information statistics}
\label{sec:2}
It is important that estimators (\ref{pdfxy},\ref{pdfxe},\ref{cpdfe}) are expressed by data and an SF that can be completely determined by repetition of the experiment. However, the basic question is: how to select a proper number of data utilized in these estimators? To answer this question we next describe the indeterminacy of the variable ${\bf z}$ by the entropy of information \cite{ig}. For this purpose we first introduce a reference PDF that is constant within the span $S_{\bf z}$: $\rho({\bf z}) = \rho(x)\rho(y)=1/(2L)^2$, and vanishes elsewhere, and define the indeterminacy of ${\bf z}$ by the negative relative information entropy \cite{ka,les,ct}:
\begin{eqnarray}
H_{\bf z}&=&-\int\!\int_{S_{\bf z}} f({\bf z}) \log \frac{f({\bf z})}{\rho({\bf z})} \,dxdy \nonumber \\
&=&\,-\int\!\int_{S_{\bf z}} f({\bf z}) \log f({\bf z}) \,dx dy - 2\log (2L).
\label{Hz}
\end{eqnarray}
By using the scattering function as the PDF, we get the uncertainty of the instrument calibration 
\begin{eqnarray}
H_{{\bf u}}&=&-\int\!\int_{S_{\bf z}} \psi({\bf z}-{\bf u})\log \psi({\bf z}-{\bf u})\,dx dy- 2\log (2L) \nonumber \\
&\approx&2\log \Bigl(\frac{\sigma}{L}\Bigr)+\log \frac{\pi}{2}+1.
\end{eqnarray}
The term $2\log (\sigma / L)$ represents the lowest attainable uncertainty of measurement. The indeterminacy $H_{\bf z}$ is generally greater than $H_{{\bf u}}$ and we define the experimental information $I(N)$ by the difference  
\begin{eqnarray}
I(N)&=&H_{\bf z}-H_{{\bf u}}\nonumber \\
&=& -\int\!\int f({\bf z}) \log f({\bf z})\,dx dy \nonumber \\
&+&\int\!\int \psi({\bf z}-{\bf u}) \log \psi({\bf z}-{\bf u}) \,dx dy .
\label{infz}
\end{eqnarray}
The quantity $I(N)$ represents the information provided by $N$ experiments on an instrument that is subject to noise \cite{ig,ct}. When sample points ${\bf z}_1, \ldots , {\bf z}_N$ are separated by several $\sigma$, the distributions $\psi({\bf z}-{\bf z}_i)$ are not overlapping and Eq.\,(\ref{infz}) yields $I(N)\approx\log N$. When distributions $\psi({\bf z}-{\bf z}_i)$ are overlapping we get $0\leq I(N)\leq \log N$. 

In an exploration the gains of measurement channels are normally set so that sample points ${\bf z}_i$ are as evenly distributed as possible over the instrument span $S_{\bf z}$.  In such a case the sample points are rather far apart when $N$ is small and yield an approximately maximal possible value $\log N$ of $I(N)$. However, with increasing $N$, the experimental information $I(N)$ increases more slowly than $\log N$ due to increasing overlapping of distributions $\psi({\bf z},{\bf z}_i)$ and therefore, measurements become ever more redundant. The difference  
\begin{equation}
R(N)=\log N -I(N)\, .
\label{Rz}
\end{equation}
thus represents the redundancy of repeated measurements in $N$ experiments. Since the overlapping of distributions $\psi({\bf z}-{\bf z}_i)$ increases with $N$, the experimental information converges to limit $I(\infty)$, and along with this, the redundancy increases logarithmically with $N$ \cite{ig}. 

The quantity
\begin{equation}
K(N)={\rm e}^{I(N)}
\end{equation}
determines the number of non--overlapping distributions that represent the experimental observation. With increasing $N$, the quantity $K(N)$ converges to a limit $K(\infty)$ that represents the complexity of the data \cite{ig,be}. It is convenient from the experimental point of view that $K(\infty)$ can be well estimated from a finite number of experiments. We could conjecture that the proper number of experiments can be specified by $K(\infty)$. However, this number can be even more conveniently estimated from the minimum of the cost function defined as the difference $C(N)=R(N)-I(N)$ of the redundancy and the experimental information. The redundancy is $R(N)=\log N -I(N)$ and hence the cost function is:
\begin{equation}
C(N)=\log N - 2I(N).
\label{Cfs}
\end{equation}
Since $I(N)$ is approximately $\log N$ at small $N$,  and is approximately constant for large $N$, the cost function $C(N)$ exhibits a minimum at a certain number $N_o$. We consider $N_o$ as the proper number of experiments to be performed for the exploration and modeling of the phenomenon. Because the redundancy is equally accounted for in the cost function as the experimental information, it turns out that the model in Eq.\,(\ref{pdfxe}) with $N_o$ data is a coarse estimator of the PDF  \cite{ig}. 

\subsection{Properties of Information Statistics}
To demonstrate the properties of statistics stemming from the information entropy we utilize the data produced by a noisy chaotic generator. This example is considered since it is often met in the exploration of complex and chaotic natural phenomena \cite{gs,mo}, and since it makes comparison feasible with the expected properties of the modeling. At the generation of data the variables $x_i$ and $y_i$ were comprised as: $x_i=x_{o,i}+n_{x,i}$ and $y_i=y_{o,i}+n_{y,i}$, where $x_{o,i}$ and $y_{o,i}$ are two successive chaotic values that are related by a logistic map \cite{gs,mo}, while terms $n_{x,i},n_{y,i}$ represent measurement noise calculated by independent random generators with zero mean and a standard deviation $\sigma=0.2$. This noise corresponds to the Gaussian SF $\psi({\bf z})={\rm g}(x,0.2){\rm g}(y,0.2)$. 

\begin{figure}
\centering
\includegraphics[width=2.5in]{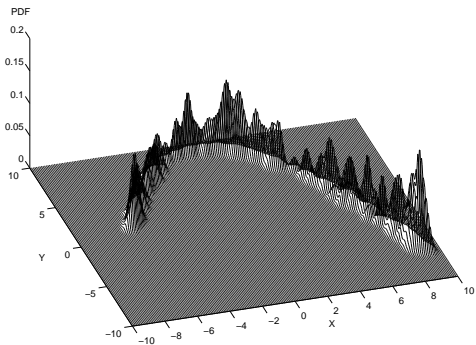}
\caption{The joint PDF $f({\bf z})$ utilized to demonstrate the properties of statistics $I,R,C$ and the conditional average estimator.}
\label{figpdfz}
\end{figure}

\begin{figure}
\centering 
\includegraphics[width=2.5in]{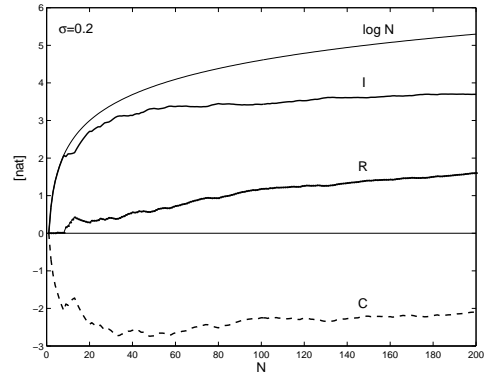}
\caption{Dependence of $\log N$, experimental information $I$, redundancy $R$, and cost function $C$ on the number of samples $N$. Statistics are expressed in the natural unit of information $nat$.}
\label{figIRC}
\end{figure}

For the demonstration we first formed the basic data set $\{x_i, y_i \}$ with $N=200$ samples. These data were used to estimate the joint PDF by Eq.\,(\ref{pdfxy}). The graph of the estimated PDF is shown in Fig.\,\ref{figpdfz}, while graphs of corresponding experimental information $I$, redundancy $R$, and cost function $C$ are shown in Fig.\,\ref{figIRC}. In the same figure the maximal possible information is presented by the curve $\log N$. 

The experimental information $I(N)$ converges with increasing $N$ to $I(\infty )\approx 3.8$ which yields $K(\infty)\approx 45$. Due to the convergence of experimental information the curve $I(N)$ starts to deviate from $\log N$ with increasing $N$. Consequently, the redundancy $R=\log N-I(N)$ starts to increase, and this leads to a minimum of the cost function  $C(N)=\log N-2I(N)$. The minimum, which occurs at $N_o\approx 32$, is not very pronounced due to statistical variations. $N_o$ is smaller but close to $K_\infty$. 

\begin{figure}
\centering 
\includegraphics[width=2.5in]{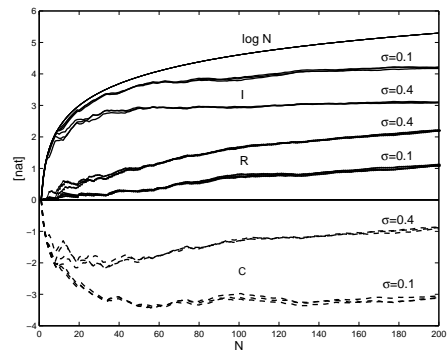}
\caption{Dependence of $\log N$, experimental information $I$, redundancy $R$, and cost function $C$ on the number of samples $N$,  determined from various data sets at $\sigma=0.1$ and $0.4$.}
\label{figIRC0104}
\end{figure}

To demonstrate the influence of scattering width and statistical variation on the presented statistics the calculations were repeated for  $\sigma=0.1$ and $0.4$ with three different sample sets. The results are shown in Fig.\,\ref{figIRC0104}. As could be expected, the limit value of $I$ increases with decreasing $\sigma$. This property is consistent with the well-known fact that more information can be obtained by using an instrument of higher accuracy, which corresponds to a lesser scattering width. In contrast to this, the redundancy of measurement decreases, and along with it, the optimal number $N_o$ increases with the decreasing scattering width.

\section{Estimation of a Physical Law}
\label{sec:5}
The example shown in Fig.\,\ref{figpdfz} resembles a ridge along a line $y_o(x)$ which we want to extract from the given data in an optimal way. For this purpose we select from a set of joint data only those that all have a certain value $x$. These joint data generally exhibit various values of $y$. We consider as an optimal predictor of the variable $y$ from a given value $x$ the value $y_p$ at which the mean square prediction error is minimal:
\begin{equation}
{\rm E} [(y_p - y)^2|x]\, = \,{\rm min}(y_p)
\end{equation}
The minimum occurs where $d {\rm E} [(y_p - y)^2|x]/dy_p=0$, which yields as the optimal predictor $y_p$ the conditional average: 
\begin{equation}\label{CA}
y_p(x)\,=\,{\rm E} [y|x]\,=\,\int_{S_y} y \,f(y|x) \,dy 
\end{equation}
By using Eq.\,(\ref{cpdfe}) we express the conditional average as:
\begin{equation}\label{CAN}
y_p(x)\,=\,\frac{\sum_{i=1}^N y_i \psi (x-x_i,\sigma)}{\sum_{j=1}^N \psi (x-x_j,\sigma)}=\sum_{i=1}^N y_i C_i (x).
\end{equation}
The coefficients 
\begin{equation}\label{C}
C_i (x)\,=\,\frac{\psi (x-x_i,\sigma)}{\sum_{j=1}^N \psi (x-x_j,\sigma)}
\end{equation}
satisfy the conditions
\begin{equation}
\sum_{i=1}^N C_i (x)=1\, ,\, 0 \leq C_i (x) \leq 1. 
\end{equation}
The coefficient $C_i (x)$ can be interpreted as a normalized measure of similarity between the given $x$ and the sample $x_i$. The calculation of $y_p(x)$ corresponds to an associative recall of memorized items, which is a property of an intelligence. Therefore, the estimator $y_p(x)$ could be treated as a basis for the development of machine intelligence based on modeling of natural laws \cite{gs,ha}.

A predictor maps the stochastic variable $x$ to a new stochastic variable $ y_p$ that generally differs from the variable $y$. When the variables $x,y$ are related by some physical law and the measurement noise is small, we expect that the first and second statistical moments ${\rm E}[ y- y_p]$, ${\rm E}[ (y- y_p)^2]$ of the prediction error are also small. The second moment is: ${\rm E}[ (y- y_p)^2]={\rm Var} (y)+{\rm Var} ( y_p)-2{\rm Cov} (y, y_p)+[{\rm m}(y)-{\rm m}( y_p)]^2$, where ${\rm E}, {\rm m}, {\rm Var},{\rm Cov}$ denote statistical average, mean value, variance and covariance, respectively. In the case of statistically independent variables $y$ and $ y_p$ with equal mean values we get: ${\rm E}[ (y- y_p)^2]={\rm Var} (y)+{\rm Var} ( y_p)$. With respect to this property we define the predictor quality by the formula
\begin{eqnarray}
Q&=&1-\frac{{\rm E}[ (y- y_p)^2]}{{\rm Var} (y)+{\rm Var} ( y_p)} \nonumber \\
&=&\frac{2{\rm Cov} (y, y_p)}{{\rm Var} (y)+{\rm Var} ( y_p)}-\frac{[{\rm m}(y)-{\rm m}( y_p)]^2}{{\rm Var} (y)+{\rm Var} ( y_p)}
\end{eqnarray}
The quality is 1 if the prediction is exact: $y_p= y$, while it is 0 if $y$ and $ y_p$ are statistically independent and have equal mean values. The quality $Q$ may be negative if ${\rm m}(y)\ne {\rm m}( y_p)$. 

For the predictor defined by the conditional average $y_p(x)\,=\,\int y \,f(y|x) \,dy$, we analytically obtain the equalities:
${\rm m}(y)={\rm m}(y_p)$ and ${\rm Cov} (y, y_p)= {\rm Var} ( y_p)$, which yield  
\begin{equation}
Q=\frac{2{\rm Var} ( y_p)}{{\rm Var} (y)+{\rm Var} ( y_p)}.
\label{QCA} 
\end {equation}
From the definition of the conditional average it follows $0\le {\rm Var} ( y_p)\le {\rm Var}(y)$ and therefore $0\le Q\le 1$. This inequality need not be fulfilled exactly if CA is statistically estimated from a finite number of samples. With increasing $N$ we generally expect that the CA statistically estimated by Eq.\,(\ref{CAN}) increasingly better represents the underlying physical law. In relation to this expectation there arises the question of whether the number $N_o$ of data yields a judicious estimation of the underlying law.

\subsection{Properties of CA Predictor}
\label{sec:6}

\begin{figure}
\centering 
\includegraphics[width=2.5in]{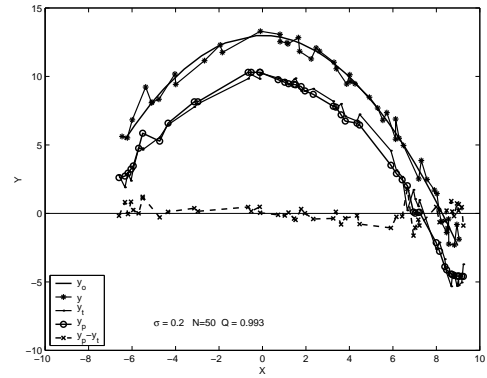}
\caption{Testing of the CA predictor. Graphs represent the underlying law $y_o$ and given data $y$ -- (top two), test $y_t$ and predicted data $y_p$ -- (middle two), and prediction error $y_p-y_t$ -- (bottom). Graphs are displaced in the vertical direction for better visualization.}
\label{figCAN}
\end{figure}

To answer the last question we demonstrate the properties of the CA predictor for the case of noise-corrupted chaotic data with standard deviation of scattering $\sigma=0.2$. From the set of $200$ data that were used when estimating the joint PDF by Eq.\,(\ref {pdfxy}), a reduced set $\{x_i, y_i ;\,i=1,\ldots,N=50\}$ was utilized for the sake of clear presentation. These basic data are shown by stars in the top curve of Fig.\,\ref{figCAN} together with the underlying law $y_o(x)$. 

The conditional average predictor was modeled by inserting data from the basic data set into Eq.\,(\ref{CAN}). To demonstrate its performance, we additionally generated a test data set $x_{i,t},y_{i,t} $ with different seeds of random generators. Using the values $x_{i,t}$ of the test set, we then calculated the corresponding values of $y_p$ by the modeled CA predictor. The test and predicted data are shown by the middle two curves in Fig.\,\ref{figCAN}. The prediction error $y_p-y_{t}$, calculated from both data sets, is presented by the bottom curve in Fig.\,\ref{figCAN}. The curve joining the predicted data is smoother than the curve joining the original test data. The smoothing is a consequence of estimating the CA from various data $y_i$ from the basic data set. In spite of this difference between both curves, we can intuitively conclude that rough properties of the hidden law $y_o(x)$ are properly revealed by the predictor. 

\begin{figure}
\centering 
\includegraphics[width=2.5in]{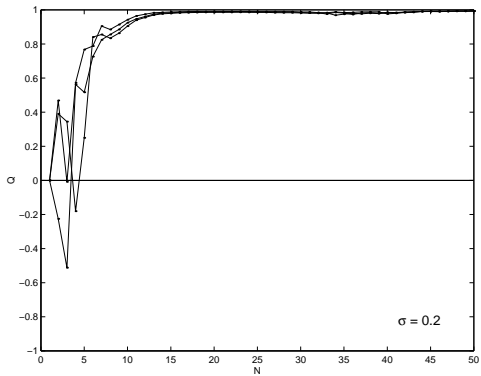}
\caption{Dependence of predictor quality $Q$ on the number of samples $N$ determined by various statistical data sets.}
\label{figCAQ}
\end{figure}

The quality of the CA predictor depends on sets of samples utilized in statistical modeling and testing. To demonstrate this dependence, we repeated the modeling and testing three times, using various statistical sample sets with increasing $N$. The estimated predictor quality $Q$ is presented in Fig.\,\ref{figCAQ} as a function of the number of samples. 
For each data set the statistical fluctuations decrease with increasing $N$, so that qualities converge to the same limit. With increasing $N$, the curves determined from different data sets merge approximately at the number $N_{CA}=15$. At the previously determined optimal number $N_o=32$ the quality is above $0.99$. The difference between curves is there about two orders of magnitude smaller than the corresponding quality and apparently disappears at $K(\infty)\approx 45$. With respect to these properties we argue that in the present case about $N_o$ data values already provides for a judicious modeling of the underlying law $y_o(x)$ by the CA predictor.

The quality of the CA predictor exhibits a convergence to some limit value that characterizes the applicability of the modeling. 
The limit value generally increases with the decreasing scattering width $\sigma$, but on average $Q$ is less than $1$ if $1/\sigma$ and $N$ are finite. This means that it is not possible to determine exactly the underlying physical law $y=y_o(x)$ based on joint data obtained by an instrument influenced by stochastic disturbances.\\

\section{Conclusion}
\label{sec:7}

Our approach indicates that the objectively introduced kernel estimator provides for a nonparametric statistical modeling of an explored phenomenon that can be automatically performed by a computer in a measurement system. Estimated statistics provide answers to basic questions about the quantity of information provided by experiments, the complexity of the data, the proper number of data needed for the modeling, and the quality of the predictor of the underlying physical law.

The estimated physical law represents the distribution of the variable $y$ at a given value $x$ by a single value $y_p(x)$. More in tune with our interpretation, the corresponding conditional PDF is mapped to the scattering function $f(y|x)\mapsto \psi(y-y_p(x))$. Such a mapping is generally accompanied by a reduction of the entropy of information that corresponds to certain information gain. This property is opposite to the loss of information caused by stochastic disturbances in signal transmission channels \cite{sha}. If the gaining of information from observations is considered as a basis of natural intelligence \cite{ha,ka}, then a system that is capable of estimating a physical law from measured data autonomously has to be treated as an intelligent unit whose level of intelligence can be quantified by the information gain. Such an interpretation provides a common basis for a unified treatment of experimental sciences and natural or artificial intelligence \cite{gs,ha}.

\section*{Acknowledgment}
This work was supported by The Ministry of Higher Eduacation, Science and Technology of Republic Slovenia and EU - COST.


\end{document}